\documentclass[12pt]{article}
\usepackage{times}
\usepackage{geometry}
\usepackage{graphicx}
\usepackage[dvipsnames]{xcolor}
\usepackage{natbib}
\usepackage{enumitem,amssymb}
\usepackage{ragged2e}
\newlist{thematic}{itemize}{8}
\setlist[thematic]{label=$\square$}
\usepackage{pifont}

\usepackage{hyperref}
\newcommand\myshade{100}
\colorlet{mylinkcolor}{violet}
\colorlet{mycitecolor}{YellowOrange}
\colorlet{myurlcolor}{Aquamarine}
\hypersetup{
  linkcolor  = mylinkcolor!\myshade!black,
  citecolor  = mycitecolor!\myshade!black,
  urlcolor   = myurlcolor!\myshade!black,
  colorlinks = true,
}
\usepackage[capitalize]{cleveref}

\begin{document}

\title{Summary of the 14th IACHEC Meeting}

\author{
K.~K.~Madsen$^a$,
Y.~Terada$^{b,c}$,
V.~Burwitz$^{i}$,
G.~Belanger$^d$,
C.~E.~Grant$^e$,
M.~Guainazzi$^f$, \and
V.~Kashyap$^g$, 
H.~L.~Marshall$^e$, 
E.~D.~Miller$^e$, 
L.~Natalucci$^h$, 
P.~P.~Plucinsky$^g$ \and
}

\date{\today}   

\newcommand{\artxc}{\textit{ART-XC}}
\newcommand{\erosita}{\textit{eROSITA}}
\newcommand{\rosat}{\textit{ROSAT}}
\newcommand{\xmm}{XMM-\textit{Newton}}
\newcommand{\chandra}{\textit{Chandra}}
\newcommand{\suzaku}{\textit{Suzaku}}
\newcommand{\swift}{\textit{Swift}}
\newcommand{\nicer}{\textit{NICER}}
\newcommand{\astrosat}{\textit{Astrosat}}
\newcommand{\nustar}{\textit{NuSTAR}}
\newcommand{\hxmt}{\textit{Insight-HXMT}}
\newcommand{\hitomi}{\textit{Hitomi}}
\newcommand{\integral}{\textit{INTEGRAL}}
\newcommand{\fermi}{\textit{Fermi}}
\newcommand{\athena}{\textit{Athena}}
\newcommand{\ep}{\textit{Einstein Probe}}
\DeclareRobustCommand{\ion}[2]{\textup{#1\,\textsc{\lowercase{#2}}}}
\newcommand{\cstat}{{\tt c-stat}}

\maketitle

{\centering
 $^a$Cahill Center for Astronomy and Astrophysics, California Institute of Technology, USA \\
  $^b$Saitama University, Japan \\
  $^c$Japan Aerospace Exploration Agency, Institute of Space and Astronautical Science, Japan \\
  $^d$ESA-ESAC, Spain \\
  $^e$Kavli Institute for Astrophysics and Space Research, Massachusetts Institute of Technology, USA \\
  $^f$ESA-ESTEC, The Netherlands \\
  $^g$Center for Astrophysics $|$ Harvard \& Smithsonian (CfA), USA \\
  $^h$IAPS-INAF, Italy
  $^i$Max Planck Institute for Extraterrestrial Physics, Germany\\
}

\vspace{5mm}
\begin{center}
{\bf \large Abstract} \\
\end{center}
We summarize the 14th meeting of the International Astronomical Consortium for High Energy Calibration (IACHEC) held at \textit{Shonan Village} (Kanagawa, Japan) in May 2019. Sixty scientists directly involved in the calibration of operational and future high-energy missions gathered during 3.5 days to discuss the status of the cross-calibration between the current international complement of X-ray observatories, and the possibilities to improve it. This summary consists of reports from the various WGs with topics ranging from the identification and characterization of standard calibration sources, multi-observatory cross-calibration campaigns, appropriate and new statistical techniques, calibration of instruments and characterization of background, communication and preservation of knowledge, and results for the benefit of the astronomical community.

\section{Introduction}

The International Astronomical Consortium for High Energy Calibration (IACHEC)\footnote{\href{http://iachec.org}{\tt http://iachec.org}} is a group dedicated to supporting the cross-calibration of the scientific payload of high energy astrophysics missions with the ultimate goal of maximizing their scientific return. Its members are drawn from instrument teams, international and national space agencies, and other scientists with an interest in calibration. Representatives of over a dozen current and future missions regularly contribute to the IACHEC activities. Support for the IACHEC in the form of travel costs for the participating members is generously provided by the home institutions and thus relevant funding agencies of participating members.

IACHEC members cooperate within Working Groups (WGs) to define calibration standards and procedures. The objective of these groups is primarily a practical one: a set of data and results produced from a coordinated and standardized analysis of high-energy reference sources that are in the end published in refereed journals. Past, present, and future high-energy missions can use these results as a calibration reference.

The 14th IACHEC meeting was hosted by Yukikatsu Terada (Saitama U./JAXA), Masahiro Tsujimoto (JAXA), Koji Mori (Miyazaki U.), Hiroshi Nakajima (Kanto Gakuin U.), Mutsumi Sugizaki (TITEC), Tatehiro Mihara (RIKEN), Takuya Midooka (JAXA/U. Tokyo), Hiromasa Suzuki (U. Tokyo), Hikari Kashimura (Kanto Gakuin U.), and Tomokage Yoneyama (Osaka University). It was held at \textit{Shonan Village} (Kanagawa, Japan) in Japan. The meeting was attended by 61 scientists from the US, Europe, and Asia, representing high energy missions based in US, UK, Italy, Germany, France, Netherlands, Switzerland, Spain, China, India, and Japan. The meeting also featured presentations from IACHEC members unable to attend in person, who were able to view and contribute to the meeting via video teleconferencing. Advances in the understanding of the calibration of more than a dozen missions was discussed, covering multiple stages of operation and development:

\begin{itemize}

\item[-] Recently completed missions - \suzaku\ and \hitomi
\item[-] Currently operating missions - \chandra, \xmm, \swift, \integral, \astrosat, \nustar, \nicer, and \hxmt\
\item[-] Pre-launch status - \erosita
\item[-] Missions under development: \athena, \ep, \textit{eXTP}, \textit{SVOM}, and \textit{XRISM}, 
\item[-] Ground calibration facilities: acceleration facilities at IHEP and MPE/PANTER
\item[-] Other: AtomDB

\end{itemize}

This year's special sessions discussed the impact of calibration uncertainties on modeling astrophysical effects, and the effects of atomic data on calibration.
This report summarizes the main results of the 14th meeting\footnote{The presentations made at the meeting are available at: \\ \href{https://iachec.org/shonan-village-center-japan-20-23-may-2019/}{\tt https://iachec.org/shonan-village-center-japan-20-23-may-2019}.}, and comprises the reports from each of the IACHEC WGs: Calibration Uncertainties (\cref{s:calstat}), Communication (\cref{s:comm}), Contamination (\cref{s:contam}), Coordinated Observations (\cref{s:coord}), Detectors and Background (\cref{s:det_bkg}), Heritage (\cref{s:heritage}), High-Resolution Spectra (\cref{s:hires}), Non-Thermal SNRs (\cref{s:nt_snr}), Thermal SNRs (\cref{s:t_snr}), Timing (\cref{s:timing}), and Isolated Neutron Stars and White Dwarfs (\cref{s:inswd}).

The IACHEC gratefully acknowledges sponsorship for the meeting from the Society for Promotion of Space Science and the International Conference Support Program of Kanagawa Prefecture Japan.

\section{Working Group reports}

\subsection{Calibration Uncertainties: CalStats}\label{s:calstat}

The CalStats WG focuses on statistical methodology as applied to calibration data and analysis. This includes documenting and recommending good analysis practices, developing techniques that are mathematically robust, and finding methods that deal with commonly encountered calibration analysis problems.

During the 14th IACHEC, there were three broad themes among presentations that were relevant to this WG.

\paragraph{1. Dealing with atomic model variances} 

Due to the proximity of the venue to \hitomi\ and the XRISM development and operations, the consequences of modeling high-spectral resolution data using imperfect knowledge of atomic physics was a popular topic. Several of the talks during the High-Resolution WG (\cref{s:hires}) dealt with this.

R.~Smith (CfA, USA) described efforts to systematically compare plasma model codes, and the tests that were developed during the Lorentz meeting;%
\footnote{\url{http://www.atomdb.org/Issues/lorentz.php}} preliminary results show substantial differences between AtomDB, Chianti, SPEX, ADAS, and SASAL codes. 

A.~Foster (CfA, USA) described efforts aimed at understanding what to do once uncertainties in atomic data become widely available, and showed the consequences of imposing reasonable uncertainties on parameters like collision strengths, Einstein coefficients, recombination rates, autoionization rates, etc. 

M. Sawada (GSFC, USA) described in detail differences between AtomDB and SPEX for both equilibrium and NEI plasma, demonstrating that flux differences as high as $2\times$ are possible in the Fe~K$\alpha$ region in certain cases.

G.~Betancourt-Martinez (IRAP, France) described examples where atomic codes were confronted with high-resolution observational data, e.g., for O\,I K$\alpha$, requiring new calibration measurements. 

J.~Kaastra (SRON, Netherlands) compared, during one of the plenary sessions, the effects of uncertainties in calibration  and in thermal atomic models, concluding that the latter can play a critical role that can, in some instances, assist in calibration. 

Lastly, J.~Garcia (Caltech, USA) described how spectral model shapes drive understanding of physical processes in accreting sources, and the resolution of several questions are limited by imperfect cross-calibration.

\paragraph{2. Modeling corrections to calibration products} 

Improvements in calibration are often the result of small empirical corrections to physically motivated approximate models. These usually involve small corrections developed by comparing predicted responses to observed data.  Three such updates were presented during the meeting as follows.

K.~Dennerl (MPE, Germany) described his continuing efforts (in plenary as well as in the White Dwarfs and Thermal SNR WG (\cref{s:t_snr}) to model the \xmm/EPIC-pn RMF via several modifier functions that tweak the profile of the probability density that describes the response of the detector at a given photon energy.  He established the parameters of the shaping functions based on a fixed astrophysical model derived from \chandra/LETG observations of RX\,J1856.3-3754 (RXJ1856), and could then successfully fit \/EPIC-pn spectra of supernova remnants 1E\,0102-72 (E0102) and N132D to significantly reduce the scatter in the flux estimates at different energies. 

C.~Pommranz (IAAT, Germany) described, during a plenary session, updates to the {\tt CORRAREA} tool (viz.\ Read et al.\ 2014) that was developed to bring \xmm/EPIC-MOS and \xmm/EPIC-pn fluxes into coherence. It is a multiplicative energy-dependent empirical correction factor applied to \xmm/EPIC-MOS effective areas to match them to \xmm/EPIC-pn based on a sample of 347 objects from 3XMM-DR7 selected to be near on-axis, isolated point sources at high Galactic latitudes with large number of counts. The residual ratio of the fluxes in different spectral ranges for various instrument modes are fit by a smooth function of nominal measured energy.  It is anticipated that this will become part of SAS in a future release. 

T.~Gaetz (CfA, USA) presented a parametric model of the particle background spectrum (see also Bartalucci et al.\ 2014) seen in both the ACIS-I and ACIS-S arrays obtained during ``stowed'' observations when the ACIS is kept on but outside the FOV.  He has developed a partially physics-based model incorporating fluorescent lines (from Al\,K$\alpha$, Si\,K$\alpha$, Au\,M$\alpha$,$\beta$, Ni\,K$\alpha$,$\beta$, Au\,L$\alpha$, and Au\,L$\beta$) from both the detector FOV and the framestore, and with and without VFAINT mode cleaning\footnote{\url{http://cxc.harvard.edu/ciao/why/aciscleanvf.html}}, and with the continuum modeled as a combination of a shallow power-law ($\Gamma\sim0.08$) and an exponential low-energy peak.  This is part of a new effort spearheaded by IACHEC to develop well-defined models for describing the background spectra so that they can be easily used as part of simultaneous source$+$background spectral fitting.

\paragraph{3. Detailed mathematical and statistical descriptions} A core purpose of the CalStats WG is to elucidate and expound accurate mathematical descriptions of observational processes.  

An example of this was the description by Y.~Terada, during a plenary session, of the process of timing calibration and the uncertainties that characterize it. He pointed out the differences between verification of timing performance and understanding uncertainties in the calibration (see also Terada et al.\ 2018) and described the effect of photon number statistics on determining the pulse profile widths and time resolution. 

During the CalStats WG session, G.~Belanger (ESAC, Spain) continued with this theme, demonstrating that timing analysis can be made significantly more precise if one were to use unbinned data, whether computing periodograms or detecting transients.  He points out that period searchers are more efficient by reducing low-frequency fluctuations when the classical Rayleigh $R_k^2$ and $Z^2$ statistics are modified to account for inter-harmonic oscillations (see Belanger 2016); the generalized Rayleigh statistic ${\cal{R}}_k^2$ and the modified ${\cal{Z}}^2 = \sum_k {\cal{R}}_k^2$ are less prone to fluctuations at low frequencies and have more power at lower S/N thresholds.  Similarly, transient detection has significantly more power when the proper likelihoods are used in computing the probability of deviations from a non-varying baseline (see Belanger 2013).  

V.~Kashyap (CfA, USA) sounded a warning about the potential misuse of high--spectral-resolution data in the low-counts regime.  He showed that when observed photons have accurately measured energies, adding up their individual fluxes to obtain the total source flux is fraught with bias when the dataset is contaminated with particle background, specifically because these background particle events are generally not distributed in energy in the same way as the source photons. This problem can be ameliorated when the data are binned and the spectrum is explicitly modeled, but this is not always feasible when the counts are sparse. One empirical method that can mitigate this flux bias is to non-parametrically compute the probability that a given observed photon comes from the background, and perform a Monte Carlo bootstrap that removes background photons prior to computing the flux.

H.~Marshall (MIT, USA) discussed the latest results from the Concordance project (Chen et al.\ 2019), which is used to obtain a shrinkage estimate of the putative corrections to effective areas of different instruments based on commonly observed sources.  The method has been updated to incorporate separate {\sl a priori} variances (the so-called $\tau$ values) on systematic uncertainties in different instruments, which were collected during previous IACHEC workshops.  The analysis was performed on the datasets of E0102 (Plucinsky et al. 2017), 2XMM (Watson et al.\ 2009), and XCAL (Stuhlinger et al.\ 2010), and showed several qualitative changes compared to previous results.  

First, because of the large differences in the adopted $\tau$ (ranging from $\sim$2\% for \xmm\ to $\sim$15\% for \suzaku), the required EA corrections based on E0102 are now predominantly driven by \xmm\ and \chandra\ results. Second, while relatively large corrections are estimated for \xmm/MOS, \suzaku, and \swift, the uncertainties on these estimates are large enough to cover the case of no required change. And third, where previously \xmm/MOS and \xmm/pn were all within no required correction for the XCAL sample, now an increase in the EA is clearly required for MOS at high energies, consistent with {\tt CORRAREA} findings (see above).  

These estimates can still be refined based on newer estimates of the flux ratios for E0102 and incorporating correlations across energy bands (which are currently neglected).  The next steps in the analysis is to estimate how much of the size of the error bars depend on the adopted values of $\tau$, and thus to explore whether they are justified, and if so, whether they indicate to the calibration scientists of the respective instruments where to focus their efforts at improving the effective areas.

\paragraph{The CalStats WG has several goals that will define its work over the following years:}
\begin{itemize}
\item[-] codifying ARF uncertainties in several more systems (including \astrosat) using the MCCal method (Drake et al.\ 2006, J.Drake et al.\ 2019 in preparation), expanding the RMF modeling method of Dennerl to ACIS, and thence to a generalized form based in pyBLoCXS (Lee et al.\ 2011, Xu et al.\ 2014); 
\item[-] bringing the state-of-the-art method developed to incorporate atomic emissivity uncertainty in solar coronal density measurements (based on pyBLoCXS; Yu et al.\ 2018) to the analysis of O\,VII, O\,VIII, and Fe\,XVII lines in the Capella X-ray spectrum;
\item[-] continuing the development of well-defined XSPEC and CIAO/Sherpa models to define the particle backgrounds in various detectors, and to also develop methods to include smooth perturbations on it (Algeri 2019) so that observation-specific modifications can be easily made within the standard fitting process; 
\item[-] upgrading the Concordance method (Chen et al.\ 2019) to account for cross-passband correlations, apply it to more datasets, and publish in an astronomy journal (H.Marshall et al.\ 2019 in preparation); and in addition, 
\item[-] we plan on compiling a set of statistical tips and tricks that will form a reference guide for the analysis of high-energy calibration data.
\end{itemize}

\subsection{Communication}\label{s:comm}

The Communication WG was created in order to improve communication across the different WGs and to increase visibility of IACHEC activities in the astrophysics community. The meeting in May 2019 was the first in which members of the WG met in person to discuss ideas for possible improvements. Here is a summary of what was discussed and decided during this meeting.

The decisions that were made in relation to communications included the migration of the IACHEC website to a commercial server on Wordpress. This was based on the need to update, modernise, and ensure independence from the institutions to which IACHEC members have affiliations. The migration to Wordpress would also allow an straight-forward handling of all email lists used by the different WGs that have until 2019 been maintained by the individual WG chairs on different servers and in different ways. In addition, hosting on Wordpress would offer an easy way to publish a newsletter or any other kind of communication with ease in a way that allows people to subscribe and unsubscribe on their own. You can visit the new website at \href{https://iachec.org}{iachec.org}.

The WG meeting was initiated with a short survey from which it was generally agreed that both internal and external communications could be improved, and that this could be done through a variety of means that included: increased presence on social media, circulation of a regular newsletter; and presentation of more talks and posters at conferences. Efforts will be made to follow up on these conclusions over the coming years.

\subsection{Contamination}\label{s:contam}

The Contamination WG exists to share information between soft X-ray instruments that suffer from molecular contamination (e.g., Marshall et al.\ 2004, Koyama et al.\ 2007, O'Dell et al.\ 2013).  Since its inception, the WG has covered three broad topics: (1) comparison of contamination among instruments and missions; (2) mitigation for current instruments; and (3) mitigation for future instruments.  The WG met for one session which included updates from several operating missions along with participation from calibration scientists of past and future missions.

H.~Marshall presented the status of the \chandra/ACIS contamination model, formulated from regular imaging observations of the galaxy cluster Abell 1795 and dispersed LETG/ACIS spectra of the blazar Mkn 421.  Recent observations show that the contaminant deposition continues at a rate that has been fairly constant for the past 3--5 years.  As noted in past meetings, the material accumulating recently has a higher O:C ratio than that accumulating in 2001--2005, with X-ray absorption fine structure in the carbon edge suggesting that it is aromatic as opposed to aliphatic, as it was for the first 5 years of contamination buildup.  
The spatial distribution of accumulated material is still asymmetric, with more along the outer perimeter of the ACIS-I and -S arrays, however the current accumulation is more spatially uniform.  The ACIS contamination model is expected to be updated later this year.

P.~Plucinsky (CfA, USA) presented observations of 1E0102.2-7219 (E0102), a thermal SNR target used to verify the contamination model due to its strong, isolated He-like and H-like O and neon lines.  This verification is done by including the contamination model (version N0012) in the ACIS response and then fitting the E0102 IACHEC model, allowing only the four line normalizations and an overall normalization to vary (see Plucinsky et al.\ 2017).  In recent ACIS-S3 observations from 2017--2018, the O\,\textsc{viii} line normalization is underestimated compared to the IACHEC model, a trend that continues in new data from 2019 and now extends to the Ne\,\textsc{ix} and Ne\,\textsc{x} normalizations and to ACIS-I3 as well, at both on-axis and off-axis locations on both chips.  This suggests that the soft-band effective area is overestimated, likely because the amount of contaminant is underestimated, consistent with the findings detailed by Marshall. Plucinsky noted that the O\,\textsc{vii} normalization shows conflicting behavior at recent times, however the uncertainty in the normalization is now large due to the low count rate.

J.~Kaastra presented work currently underway to produce self-consistent, time- and wavelength-dependent correction factors to the \xmm/RGS effective area, including the effects of contamination.  This study uses $\sim$100 spectra of the blazars Mkn 421 and PKS 2155-304 taken with RGS 1 and 2 over the course of the \xmm\ mission.  Around the nitrogen edge, there is a strong, time-dependent residual feature that is consistent with adsorption of ammonia on the Al$_2$O$_3$ surface of the RGS.  The ammonia itself likely arises from the hydrazine (N$_2$H$_4$) propellant used on the spacecraft.  Near the O edge, work is underway to collect a number of effects, including contamination, into a single correction factor.  Measurement of the O component in the contaminant is complicated by the neutral O in the Galactic column; after accounting for this and other effects, Kaastra showed an increase of 20\% in the neutral O column over the course of the mission, a 3.5\,$\sigma$ result.  This result is consistent between the two blazars, and indicates a build-up of O on the RGS along with the ammonia.

T.~Kohmura (TUS, Japan) presented measurements of the X-ray, UV, and optical transmission of the Contamination Blocking Filter (CBF) aboard the \hitomi/SXI.  While not a measurement of molecular contaminant itself, this work highlights an important contribution of the WG: methods to mitigate contamination while maintaining the desired instrument performance, in this case soft X-ray sensitivity.  Unlike the free-standing filters on \chandra/ACIS and \suzaku/XIS, both of which have suffered from contamination, the SXI CBF was physically separated from the cold focal plane and designed to be kept at $+$20\,C to prevent condensation of contaminants.  It initially consisted of a metal mesh with 200\,nm of polyimide, primarily for UV light blocking, coated on one surface with 30 nm of Al to protect the plastic from oxidation and provide additional optical light blocking power to the Al Optical Blocking Layer (OBL) directly deposited on the CCDs.  During ground testing, pinholes were discovered in the OBL, and the CBF was redesigned to provide additional optical light blocking, with a thicker 80 nm of Al on one side and 40 nm on the other (Tanaka et al.\ 2018).  Kohmura showed results from ground measurements of this flight model (FM) CBF, confirming that the UV and optical light blocking met requirements and that the redesigned FM CBF had reduced transmission at low energies due to the increased Al thickness.  The SXI FM CBF design will be adopted for the Xtend instrument on the upcoming XRISM mission (Tashiro et al.\ 2018).

The WG session concluded with a general discussion of future activities, including a plan for the coming year.  The WG will prepare a manuscript summarizing the work of different missions to monitor and calibrate contamination.  This manuscript will be submitted to a refereed journal, with the hope that it will provide a useful resource for future missions.

\subsection{Coordinated Observations}\label{s:coord}

The objectives of the IACHEC Coordinated Observations WG are to coordinate new observations jointly among different telescopes, analyze those observations, and publish the results.  For the 2019 IACHEC meeting, there were two presentations on cross-calibration projects and significant discussion of potential new observations.

M.~Smith (ESAC, Spain) reported results from the \xmm/\chandra\ cross-calibration effort using blazars such as 3C 273, PKS 2155$-$304, and 1H 1426$+$428.  The method is similar to that used by Ishida et al.\ (2011), where joint time intervals are constructed to make spectra for the \xmm \ (pn, MOS, and RGS) and \chandra\ \ Low and High Energy Transmission Grating Spectrograph (LETGS and HETGS) instruments.  The spectra were fit to simple power laws over small energy intervals: 0.15$-$0.33 keV, 0.33$-$0.54 keV, 0.54$-$0.85 keV, 0.80$-$1.20 keV, 1.20$-$1.50 keV, 1.50$-$1.82 keV, 1.82$-$2.20 keV, 2.20$-$3.50 keV, 3.50$-$5.50 keV, and 5.50$-$10.0 keV.  Fluxes were determined for each bandpass, which are relatively robust to the actual spectral shape and spectral index in each band, and normalized to that determined from the \xmm/pn.  Standard processing is used for all instruments, with cores excised when pileup is an issue.  The new PSF calibration provides better agreement between the pn and MOS fluxes above 3 keV, as expected.  Generally, the RGS fluxes agree well with those from the pn in each energy band in the 0.33--2.2 keV range, the MOS fluxes agree with each other but are higher by 3--10\% than those of the pn over the 0.33--10 keV range, and the \chandra\ fluxes are 5--15\% higher than the pn values over the 0.54--5.5 keV range.  Smith also looked at the consistency of fluxes when using the $\chi^2$ statistic {\tt chi} versus the Poisson likelihood-based statistic {\tt cstat}, showing clear biases when using the  $\chi^2$ statistic.

K.~Madsen (Caltech, USA) reported results from work on joint \nustar\ and \swift/XRT (WT mode) observations of bright Galactic X-ray binaries, extending the analysis to sources with different column densities and fluxes.  The best results come for sources with low fluxes and low column absorption densities, such as Her X-1 and MAXI J1820$+$070 (low state).  High column densities come with dust halos that can cause an issue if the extraction regions differ and if the core has to be excised when the source is bright.  While more work is needed to examine the effects of extraction region, it was tentatively concluded that grade 0 events are required, and that there may be a 5\% calibration issue in NuSTAR at low energies. Madsen also showed that \nustar\ and \integral/SPI spectra of MAXI J1820$+$070 agree very well over the entire \nustar\ band.

F.~F\"urst (ESAC, Spain) showed joint fits to \nustar and \xmm\ pn spectra of 3C 273, obtaining systematically spectral slopes from \nustar\ that are larger by 0.10 than for the pn, regardless of the spectral slope observed by either instrument.  By contrast, estimated fluxes in the 3--10 keV band agreed to a few percent for 3 of 5 observations.  For a sample of 10 blazars and AGN, he showed results from work by Amy Joyce, showing fluxes that are 15\% higher for \nustar\ in the 3--10 keV band than those obtained from fits to the pn data.

J.~Kennea (PSU, USA) reported on \swift\ operations, noting that the \swift project receives about 5 ToO requests and observes 60--70 different targets per day.  About 12\% of Swift observations are performed simultaneously with other missions and an additional 24\% are ToOs.  In fact, it is now standard to obtain at least 2 ks for each \nustar\ target, except when the target is observed simultaneously with \chandra\ or \xmm\ already.

Coming up in the next year is another campaign on 3C 273 that will involve \nicer, \chandra, \xmm, and \nustar\ at a minimum.  V.~Burwitz (MPE, Germany) will be coordinating observations with calibration targets to be observed by \erosita\ after its upcoming launch.

\subsection{Detectors and Background}\label{s:det_bkg}

The Detectors and Background WGs had one session each, both well-attended.  Detectors WG provides a forum for cross-mission discussion and comparison of detector-specific modeling and calibration issues, while the Background WG provides the same for measuring and modeling instrument backgrounds in the spatial, spectral and temporal dimensions. Attendees represented many past, current, and future X-ray missions, including \athena, \chandra, \ep, \hitomi, \nustar, \suzaku, and \xmm.  As existing missions go deeper, and planned missions get more ambitious, understanding and modeling background and detector response is all that much more important.

Two themes were apparent in the Detectors session, both related to the long lifetimes of many missions: long term trends in gain and response, and getting the details right.  I.~Valtchanov (ESAC, Spain) discussed the \xmm\ EPIC-pn long term CTI correction for small and large window mode, and how it differs from full frame mode.  N.~Durham (CfA, USA) gave an update on \chandra\ ACIS gain challenges, particularly warm focal plane temperatures, while T.~Gaetz showed progress on correcting the ACIS gain droop in the central columns.  H.~Miyasaka (Caltech, USA) presented improvements in the low energy response calibration and the long-term gain monitoring for \nustar.  K.~Hayashida (Osaka U., Japan) discussed a solution to the Si-K edge problem in the \suzaku\ XIS and the implications for XRISM Xtend. Finally, H.~Murakami (Tohoku Gakuin U., Japan) presented a novel clocking scheme, panning mode, to reduce pileup.

The Background session included simulations of background for future missions and measurements and modeling of in-orbit background.  S.~Stever (U. Tokyo, Japan) showed simulations of thermal excursions from cosmic rays on the \athena\ X-IFU which do not seem to present significant degradation to energy resolution, but more work is required.  Background simulations for the \ep\ Wide-field X-ray Telescope were presented by D.~Zhao (NAOC-CAS, China) including contributions from X-rays and particles and the impact of detector thickness.  E.~Miller (MIT, USA) reported on the effort to characterize and reduce the background on the \athena\ WFI.  H. Suzuki (U. Tokyo, Japan) compared Monte-Carlo simulations of atmospheric neutron and radioactivation background to the measured background on \hitomi\ HXI.  Finally, T.~Gaetz gave a status update on ACIS background modeling.

\subsection{Heritage}\label{s:heritage}

The Heritage WG has the scope of preserving the IACHEC corpus of knowledge, know-how, and best-practices for the benefit of future missions and the community at large. This is done by

\begin{itemize}
\item providing a platform for the discussion of experiences coming from operational missions;
\item facilitating the usage of good practices for the management of pre- and in-flight calibration
data and procedures, and the maintenance and propagation of systematic uncertainties (the
latter task in strict collaboration with the Systematic uncertainties IACHEC WG);
\item documenting the best practices in analyzing high-energy astronomical data as a reference for
the whole scientific community;
\item ensuring the usage of homogeneous data analysis procedures across the IACHEC calibration
and cross-calibration activities; and
\item consolidating and disseminate the experience of operational missions on the optimal calibration
sources for each specific calibration goal.
\end{itemize}

The main current activity of the WG is the development of the IACHEC Source Database (ISD).
The ISD is defined as the single repository of high-level scientific data and data analysis procedures
used in IACHEC published papers. The ISD shall be populated by an IACHEC WG whenever

\begin{itemize}
\item an IACHEC paper is published in a refereed journal; or 
\item an updated calibration relating to a published paper with a significant impact on the results of the cross-calibration analysis, as verified
by the WG, and documented in a Technical Note or in a new paper is subsequently published. 
\end{itemize}

Papers and Technical Notes should also be ingested in the ISD in addition to being made available
from the IACHEC web portal. 

The implementation of the ISD was made possible thanks to the support by the AHEAD Project funded by the European Union.

\subsection{High-Resolution Spectra}\label{s:hires}

The High-Resolution Spectra Working Group (HRWG) aims to improve interpretation of spectra from high-resolution instruments. This year, the group focused on presenting the effects of atomic data uncertainties on the modeled spectra and the resulting inferred plasma properties.

R.~Smith (CfA, USA) presented a cross-calibration effort, the Lorentz Project, which aims to provide easily cross-comparable data between different atomic database projects, including AtomDB, SPEX, CHIANTI, ADAS, XSTAR and others. This showed that there are significant areas of both agreement and disagreement, and identified that some of the disagreement is deliberate (a conscious choice to use a different model or data), while others were simply errors. This project is ongoing.

A.~Foster presented a new framework for varying the atomic data in AtomDB and running Monte-Carlo simulations to obtain numerous sets of possible plasma emissivities. Fitting the resulting spectra with these can then provide primitive estimates of the uncertainties in each of the plasma parameters when a user fits their spectrum, and can provide a reasonable measure of the robustness of the data to the underlying material. This is still preliminary work.

T.~Kallman (NASA GSFC) presented the atomic data needs project for \textit{XRISM} that aims to identify which of the mission's science goals are limited by our current knowledge of atomic data to guide future developments in experiment and theory for atomic and plasma physics and improve the mission's science return.

M.~Sawada presented the effects of changing atomic data and plasma emission models. This was performed by comparing the same spectra for a range of models, generated in SPEX and then fitted using AtomDB/XSPEC (and vice versa). The differences between the results were evident, in particular for the non-equilibrium cases. Effects on individual results should have tools developed to help users identify the causes.

G.~Betancourt-Martinez presented an overview of the significant advances in laboratory astrophysics, particularly relating to wavelengths and charge exchange models, and how they provide insight that cannot be obtained through theoretical measurements alone. 

R.~Cumbee (GSFC, USA) presented the new charge-exchange database, Kronos, created at the University of Georgia which provides detailed state selective calculation for charge exchange modeling, complete with integration into XSPEC and SPEX.

For the future, the HRWG has decided to reassess its goals. Previously, the aim was to look into a complete line list for \chandra\ spectra, starting with that of Capella. However, this is not feasible due to manpower limitations. Instead, future work will focus on two more distinct goals: (1) providing advice to future missions on pitfalls of high resolution spectroscopy, and (2) looking into measuring and calibrating responses of high resolution instruments in crowded spectra.

\subsection{Non-Thermal SNRs}\label{s:nt_snr}

The purpose of the non-thermal SNR WG is to define the two non-thermal standard candles:  G21.5-0.9 (mainly below 10 keV) and of the Crab (mainly above 10 keV) spectra. 

E.~Jourdain (IRAP, France) presented the long term spectral evolution of the Crab nebula as seen by \integral/SPI. The conclusion is that there is excellent agreement on the relative flux variations between the SPI, \fermi/GBM, and \swift/BAT. Using four different epochs the spectra were fitted with a GRBM, band model, which is a two power-law component model with a soft break between them, and a broken power-law. The GRBM model provide a significantly better fit. With this fit the slope of the Crab in the SPI is $\Gamma=-2.0$ for X-rays and $\Gamma=-2.26$ for $\gamma$-rays. The flux stability is at a few \% over 16 years and the spectral shape unchanged to first order, which means it remains an excellent standard candle for hard X-rays.

K.~Madsen presented the update on the \nustar\ Crab monitoring campaign. The campaign serves two purposes: (1) calibration, and (2) long-term monitoring of the Crab spectrum with the precision of a focusing instrument. This is accomplished by focused observations accompanied by stray-light observations as shown in Madsen et al (2017b). The campaign started in early 2016 and has collected $\sim$20 observations, obtained roughly every other month. The stray-light observations accurately track the flux, with an estimated absolute precision of 4\% and over the length of the campaign the flux has varied by less than 2\% peak-to-peak. There is an offset between the photon index measured by the focused observation and the stray-light observation of $\Delta\Gamma$ =0.01, and this off-set is well understood since the \nustar\ mirror response was calibrated against a presumed spectral slope. The relative variation of the photon index is $\sim$1\% for both stray-light and focused observations. In the next effective area update, the \nustar\ responses will be updated with the average stray-light flux and photon index.

The WG discussed updating the spectral model for both sources to the GRBM band model, since both sources exhibit a spectral break. For the Crab it is located at around 100\,keV with a gradual curvature, whereas for G21.5-0.9 this break is sharp and located at 8\,keV (Nynka et al 2013, Tsujimoto et al 2017). It is unclear if \nustar\ can sense the curvature in the stray-light X-ray spectrum of the Crab and it was decided that Madsen would apply the GBRM model to the \nustar\ Crab data to quantify the level of curvature below 80\,keV. It was also decided to apply the same model to G21.5-0.9 using data from \chandra\ (provided by N.~Durham), \hitomi\ (M.~Tsujimoto), \integral\ (V.~Savchenko), \nustar\ (K.~Madsen), \suzaku\ (TBD), \swift\ (TBD), and \xmm\ (F.~F\"urst). The data will be compiled and fit by M.~Tsujimoto.

\subsection{Thermal SNRs}\label{s:t_snr}\label{s:n132d}

The thermal SNRs WG met in one session at this IACHEC meeting. In attendance were: Andy Beardmore (U. of Leicester, UK), Konrad Dennerl, Gulab Dewangan (IUCAA, India), Nick Durham, Adam Foster, Felix F\"urst, Terry Gaetz, Kenji Hamaguchi (GSFC, USA), Eric Miller, Hiromasa Miyasaka, Hiroshi Nakajima, Paul Plucinsky, Martin Stuhlinger (ESAC, Spain), Hiromasa Suzuki, Masahiro Tsujimoto,  and Brian Williams (GSFC, USA), while Michael Freyberg (MPE, GER) and Brian Grefenstette (Caltech, US) participated remotely.  

The objectives of the thermal SNRs WG are to develop standard spectral models for 1E 0102.2$-$7219 (E0102) and N132D and to examine the calibration of the respective instruments using these spectral models and sound statistical methods.  The group wants to make more rapid progress on the IACHEC model for N132D in the coming year.  Therefore, the group decided to meet roughly once every six weeks to ensure progress is made.

\paragraph{1. 1E~0102.2$-$7219}

K.~Dennerl presented the E0102 line fluxes with his revised response matrix for the {\it XMM-Newton} EPIC~pn instrument. The revised E0102 line fluxes are now in better agreement with the {\it XMM-Newton} EPIC~MOS and {\it Chandra} ACIS.  In particular, the {Ne}~{\small
{IX}\/}~He$\alpha$~triplet and {Ne}~{\small{X}\/}~Ly$\alpha$ line agree better with MOS and ACIS, to better than 10\%. The agreement at Ne for the {\it Suzaku} XIS and {\it Swift} XRT are also better than 10\%.  The situation is not as good at the O lines. The O line fluxes still show discrepancies larger than 10\% and in some cases as large as 20\%.

M.~Stuhlinger plans to re-analyze the \xmm\ RGS data of E0102 with the latest response files for the  RGS. This is in response to an observation from J.~Kaastra that the values for the RGS effective areas have changed even for times early in the \xmm\ mission.  The expectation is that the changes to the effective areas early in the mission will be small, but since those RGS data were used extensively to develop the IACHEC model for E0102 it is important to verify this. We do not believe this affects our analysis in the IACHEC paper on E0102  (Plucinsky et al. 2017) because we allowed the normalizations of the {O}~{\small{VII}\/}~He$\alpha$~triplet,
the {O}~{\small{VIII}\/}~Ly$\alpha$ line, the {Ne}~{\small{IX}\/}~He$\alpha$~triplet, and the {Ne}~{\small{X}\/}~Ly$\alpha$ line to vary in addition to a global normalization.  We expect that the normalizations for the RGS data will change but this should not affect the other instruments unless these different normalizations argue for a change in the model.

P.~Plucinsky presented the E0102 line fluxes from the most recent \chandra\ ACIS observations of E0102 in March 2019.  These data indicate that the contamination model under-estimates the contamination in 2019 and 2020.  The \chandra\ X-ray Center calibration team expects to release an updated contamination model in Fall 2019 that will retain the effective areas for early in the \chandra\ mission but will increase the contamination at late times.

\paragraph{2. N132D}

K.~Dennerl presented fits to the N132D data with the small window mode data from the EPIC pn.  Somewhat surprisingly the fits with the current IACHEC model for the observations in which the remnant is near the edge of the window and some of the remnant is cut off are better compared to the observations in which the remnant is centered in the window.  This is puzzling because no such positional dependence is observed in the E0102 spectral fits in small window mode with the revised response matrix.  Dennerl will investigate if the background subtraction is part of the explanation. Dennerl notes that he only fits up to 3.0 keV because the new matrix was developed with E0102 and RXJ 1856 data which are both rather soft
  
 E.~Miller presented fits to the {\it Suzaku} XIS spectra of N132D. He discussed some of the issues related to the modeling of the background, in particular the model he is using for the ``Non X-ray Background (NXB)'' for the XIS.  Miller showed a zoom of the 3--11~keV region with the source and background components labelled clearly.  This figures shows how important it is to model the background carefully in order to determine an accurate line flux in the Fe~K region.  There are instrumental lines in this energy range that must be modeled well in order to determine an accurate flux.  Miller showed that with this background model he derives consistent line fluxes for the Fe~{\small{XXV}\/} and Fe~{\small{XXVI}\/} line complexes from 11 observations spread over a 5 year time period. He has a power-law component in the model for the CXB.  The normalization of this component is higher than expected for the CXB.  The suggestion from B. Grefenstette ({\it NuSTAR}) below is that there is an additional high temperature component from N132D itself.  Miller uses a dummy, diagonal response file for the instrumental background model.  This is different than what the other groups are doing.

B.~Grefenstette discussed the {\it NuSTAR} observations of N132D and the issues with background modeling with {\it NuSTAR}.  He made use of the optimal binning method described by Kaastra and Bleeker 2016. He fit with the IACHEC model and there was a clear excess at high energies that peaks around 9 or 10 keV.  He argues that an additional thermal component with a temperature around 6.5 keV fits the data
well. A similar conclusion was reached by Bamba et al. 2018 in a joint analysis of  {\it Suzaku} and {\it NuSTAR} data. Some experimentation will be done with new high-temperature component in the IACHEC model.

\paragraph{3. Cassiopeia A}\label{s:casa}

A.~Beardmore described his model for Cassiopeia~A (Cas~A) that he has been using for the {\it Swift} XRT gain calibration.  The phenomenological model, derived from  an \xmm\/ MOS1 observation of Cas~A, requires an energy scale offset of 12 eV at the Si-K$\alpha$ line when applied to the simultaneously obtained pn data. G.~Dewangan said that {\it ASTROSAT} is using this model in their calibration.  The group discussed if this model should be released as an IACHEC model.  Plucinsky noted that it is different from the E0102 and N132D models in that those models are based on high resolution gratings data while this model is based solely on CCD resolution data.  After the meeting, Andy suggested
that this model could be released as an A\&A Research Note.

\subsection{Timing}\label{s:timing}


 The Timing WG aims to provide a forum for in-orbit and on-ground timing-calibrations of X-ray missions, focusing on their timing systems, calibration methods, issues, and lessons learned. The WG also aims to coordinate simultaneous observations for timing calibrations with multi X-ray missions and/or radio observatories. On the 14th IACHEC meeting, the Timing WG was held in one parallel session with two presentations.  

 M.~Bachetti (INAF, IT) presented the improvement of timing capability of \nustar\/ mission from the design value on the timing accuracy of 100 msec in absolute timing. They corrected the frequency drifts of the quartz by the temperature and calibrated them in each ground contact, and the $\sim$20 $\mu$s stability per day (corresponds to $10^{-10}$s/s) was successfully achieved. The correction method itself is the same as \suzaku\ (Terada et al 2008) or \hitomi\ (Terada et al 2018) and is also valid and commonly useful for other missions without GPS receiver. 

 M.~Sawada presented the details of timing calibrations of the micro-calorimeter systems onboard \hitomi\ and XRISM missions. The absolute timing capabilities are already summarized in the previous 12th and 13th IACHEC meetings (Terada et al 2018) and therefore, his presentation is mainly concentrated on reporting the timing issues on the relative timing of the micro-calorimeter pixels and anti-coincidence detectors, such as threshold, grade, and pulse-height time-dependencies. 

 The Timing WG continues working to summarize the latest timing-calibration status and lessons of \chandra, \xmm, \integral, \nustar, \nicer, \hxmt, \astrosat, \erosita, \textit{XRISM} etc. In addition, the WG members plan to search for archive coordinated observations (including those with non-calibration purpose) and analyze them for timing calibration purpose. The systematic studies of dead time, grade selection, etc.\ on the timing products are also the current scope of the Timing WG. 

\subsection{Isolated Neutron Stars and White Dwarfs}\label{s:inswd}

This WG aims to improve the cross-calibration of X-ray telescopes in the low energy range ($<$ 1.5 keV) by using spectra of Isolated Neutron Stars (INSs) and White Dwarfs (WDs). These objects should not display time dependent variation and have physically well modeled spectra that can be used as spectral standard candles at low energies. Over the years, a set of white dwarfs (GD153, Sirius B and HZ43) with spectra that can be described by physical white dwarf models, and the isolated neutron star RX\,J1856.5$-$3754 with a spectrum that can be best be described by a single black-body model, have met the characteristics of a standard candle. The three WDs and RX\,J1856.5$-$3754 were used to improve the calibration of the low energy end of the \chandra\ LETGS and provide a cross-calibration with \rosat\ and recently also with \nicer. Other INSs, such as RX\,J0720.4$-$3125, looked in the beginning like a promising candidate but turned out to be variable on timescales of years.

V.~Burwitz gave an overview of the status of the INSs and WDs working Group. The status of the \chandra/LETGS observations of RX\,J1856.5$-$3754 was shown and upcoming observations in June/July 2019 in the context of \erosita \ where mentioned. All these observations will span a period of about 20 years. \nicer \ results were presented on behalf of C.~B.~Markwardt, and the RX\,J1856.5$-$3754 calibration was summarised in a talk by K.~Hamaguchi on the status of \nicer. The \nicer \ team showed that the low energy end of the spectrum could be well fit by the current standard candle single black-body model spectrum, but that at energies $>$0.5 keV a further component is required to fit the spectrum. A possible explanation for this additional component could be a contribution by the $3/4$ keV sky background.

T.~Yoneyama presented work on the universal detection of high-temperature emission in INSs. The excess detected in RX\,J1856.5$-$3754 with \nicer \ is also seen in \xmm/EPIC-pn and \suzaku\ data whilst using current calibration data. The group also looked at the spectra of other isolated neutron stars and sees evidence for a similar excess in their X-ray spectra. The true cause for this excess still needs to be understood.
 
K.~Dennerl spoke about reaching consistency between the LETGS and the \xmm/EPIC-pn using RX\,J1856.5$-$3754. For this he used a parametric EPIC-pn response function and was able to improve the low energy spectral fitting dramatically. This is still work in progress, but these optimized parameters also improve the fits to other calibration sources. 

In summary, the ISN RX\,J1856.5$-$3754 is being observed and used by all X-ray observatories for calibration purposes by monitoring the status of their low energy calibration. Work on WDs was not discussed at this year's meeting.

\section*{References\footnote{see {\tt https://iachec.org/papers/} for a complete list of IACHEC papers}}

\noindent
Algeri, S., 2019, to be submitted to Phys.Rev.D, arXiv:1906.06615 \\
\noindent
Bamba et al.\ 2018, ApJ, 854, 71\\
\noindent
Bartalucci, I., Mazzotta, P., Bourdin, H., and Vikhlinin, A., 2014, A\&A, 566, A25\\
\noindent
Belanger, G., 2013, \href{https://arxiv.org/pdf/1303.7408.pdf}{ApJ, 773, 66}\\
\noindent
Belanger, G., 2016, \href{https://arxiv.org/pdf/1712.00734.pdf}{ApJ, 822, 14}\\
\noindent
Chen Y., et al., 2019, \href{https://doi.org/10.1080/01621459.2018.1528978}{J.Am.Stat.Assoc., 114:527, 1018}\\ 
\noindent
Drake, J.J., et al. 2006, Proc. SPIE, 6270, 62701I\\
\noindent
Hitomi Collaboration et al. 2018, PASJ, 70, 16\\
\noindent
Ishida M., et al.\ 2011, PASJ, 63, 657\\
\noindent
Kaastra J. \& Bleeker J.A.M., 2016, A\&A, 587, 151 \\
\noindent
Kaastra, J.S.\ 2017, A\&A, 605, A51\\
\noindent
Koyama, K., et al.\ 2007, PASJ, 59, 23\\
\noindent
Lee, H., et al.\ 2011, ApJ, 794, 97\\
\noindent
Madsen K., et al.\ 2017a, AJ, 153, 2 \\
\noindent
Madsen K., et al.\ 2017b, AJ, 841, 5 \\
\noindent
Marshall H., et al.\ 2004, Proc.\ SPIE, 5165, 497 \\
\noindent
Nevalainen, J., et al.\ 2010, A\&A, 523, A22\\
\noindent
O'Dell S., et al.\ 2013, Proc.\ SPIE, 8559 \\
\noindent
Plucinsky P., et al.\ 2016, Proc.\ SPIE, 9905, 44 \\
\noindent
Plucinsky P., et al.\ 2017, A\&A, 597, A35 \\
\noindent
Plucinsky P., et al.\ 2018, Proc.\ SPIE, 10699, 106996B\\
\noindent
Rasmussen, C.E.\ and Williams, C.K.I. 2006, {\sl \href{http://www.gaussianprocess.org/gpml/chapters/}{Gaussian Processes for Machine Learning}}, MIT Press, ISBN 026218253X\\
\noindent
Read, A.M., Guainazzi, M., and Sembay, S., 2014, A\&A, 564, A75\\
\noindent
Stuhlinger, M., et al., 2010, \href{http://xmm2.esac.esa.int/docs/documents/CAL-TN-0052.ps.gz}{XMM-SOC-CAL-TN-0052 6.0}\\
\noindent
Tanaka, T., et al.\ 2018, JATIS, 4, 011211\\
\noindent
Tashiro, M., et al.\ 2018, Proc.\ SPIE, 10699, 1069922\\
\noindent
Terada Y., et al., 2008, PASJ, 60, 25\\
\noindent
Terada, Y., et al., 2018, \href{https://doi.org/10.1117/1.JATIS.4.1.011206}{J.Astron.Tel.Instr.Sys.\ (JATIS), 4(1), 011206}\\ 
\noindent
Terada Y., et al., 2018, JATIS, 011206\\
\noindent
Watson, M.G., et al., 2009, A\&A, 493, 339\\
\noindent
Xu, J., et al.\ 2014, ApJ, 794, 97\\
\noindent
Yu, X., et al.\ 2018, ApJ, 866, 146\\

\end{document}